\documentclass{roajarticle}
\usepackage{graphicx}
\usepackage[section]{placeins}
\usepackage{mynatbib}
\usepackage{url}
\usepackage{upgreek}
\usepackage{rotating}
\usepackage{lscape}
\newcommand{\volume}{\textbf{25}}
\newcommand{\numarul}{2}
\newcommand{\volyear}{2015}
\title{A new approach to detect the boundaries of interplanetary events}
\newcommand{\shorttitle}{A new approach to detect the boundaries of interplanetary events}
\author{Cristiana  \MakeTextUppercase{Dumitrache}}
\author{Nedelia Antonia  \MakeTextUppercase{Popescu }}
\affil{Astronomical Institute of Romanian Academy\\%
Str. Cutitul de Argint 5, 40557 Bucharest, Romania\\%
Email:  nedelia@aira.astro.ro, crisd@aira.astro.ro}

\begin{document}
\maketitle

\begin{abstract}
  We propose a new method to scan the data and then to infer the boundaries of the interplanetary coronal mass ejections, especially the interplanetary events observed by \emph{Ulysses}.
  The local minima of the temperatures ratio measured by the spacecraft are used to scan and compute the potential boundaries of the interplanetary events.
  The low plasma beta values are then invoked to detect at least four boundaries, two for the beginning and two for the end of an interplanetary coronal mass ejection (ICME). Intermediate boundaries can be identified, as indicated by other plasma and magnetic field signatures, and these ones mark substructures of a complex event.
  Using the algorithm described in this article, we have compiled a list of the ICME events registered by \emph{Ulysses} spacecraft during 2000-2002, with  their boundaries.
  After a statistical analysis, four relationships of the plasma beta with the speed are then inferred for the computed ICMEs, for each of the four detected boundaries in our compiled list.
  In order to exemplify the method we analyse in detail four events.
  This method provides premises for an alternative way of semi-automatic detection of the interplanetary events' boundaries observed by \emph{Ulysses} spacecraft.
  We introduce here also a method for visualisation of the magnetic field components that allows detection of plasma insulations and boundaries detection for other satellites too.
\end{abstract}

\section{Introduction}
\label{s1}
Coronal mass ejections (CMEs) are the largest eruptive phenomena known in the solar system as responsible for drastic changes of the space
weather conditions and directly affecting the life on Earth.
The interplanetary CMEs counterparts expand from the Sun to heliosphere and produce disturbances of the interplanetary space.
In-situ measurements of plasma and magnetic field display a wide range of signatures that guide us to identify the
interplanetary coronal mass ejections (ICMEs) and their delimitations.
Of these, widely invoked are: the intense magnetic field, the low plasma beta, density enhancements,
increasing of ion charge states and abundance of various elements, as well as a decrease of the proton temperature.
These properties are strengthened in the magnetic clouds (MCs),  events that display a distinct morphology, a magnetic-flux-rope-like structure, often accompanied by forward and reverse shocks flanking the cloud core as \emph{Ulysses} observations revealed \citep{Burlaga1991, Reley2007}, magnetic field smooth rotation, and bi-directional electron streaming \citep{Crooker2004}.
MCs, representing a subset of interplanetary ejecta, are the interplanetary manifestation of a flux rope expelled from the Sun, and
were first defined by \citet{KleinBurlaga82} as regions with a radial dimension of approximately 0.25 AU at 1 AU, inside which the magnetic field strength is high.
\citet{Wei2003JGRA} analyzed more magnetic clouds at 1 AU and found distinct characteristics of plasma and magnetic fields inside the magnetic cloud bodies compared to the boundary layers.
In addition, we point-out as an useful tool to detect the presence of an MC and its boundaries, the minimum variance analysis \citep{KleinBurlaga82,BothmerSchwenn98,Dumi2011}, that proves a coherent rotation of the MCs around one axis, at the spacecraft encounter \citep{Burlaga1988}.
This tool often indicate the existence of substructures as merged clouds, beside the function of refining the MCs' boundaries.
In contrast to MC, the non-MC ICMEs could have weaker magnetic fields, higher proton temperatures, higher plasma beta, when comparing to the MCs, as \citet{Burlaga2001} reported. These ICMEs can be the result of the interaction of two or more MCs at lower heliocentric distances -- they were called  \textit{complex ejecta} ICMEs (non-MCs).

Many times, the ICMEs or MCs are merged, even if they left the Sun as distinct events.
In this case,  it is a challenge to detect the real boundaries of each one or how much two events are mixed, or to
understand the mixed and non-mixed parts of a complex phenomenon.
The solar source identification and analysis are useful in this case. But, \citet{Richardson2010} have shown that it is possible to not really find the solar source for $46\%$ of Earth directed ICMEs.
In a statistical survey of the ICMEs observed near-Earth, \citet{Richardson2004grl} showed that the MCs fraction related to the total number of the ICMEs depends on the solar cycle.
In the list built by \citet{Du2010}, with 181 ICMEs observed by \emph{Ulysses} spacecraft, only $43\%$  of events were identified as MCs.

Beyond one astronomical unit (AU), the ICME identification become more difficult because some ICME signatures are blurred through the interaction with the ambient solar wind.
\emph{Ulysses} spacecraft, launched in 1990, was the first mission that explored the Heliosphere from the solar equator to the poles.
Its observations for these long distances, as well as those obtained during its transit over the solar poles, provided us important information and an understanding on how the high latitude CMEs travel through space; it gave us an insight to the 3D heliosphere.
\emph{Ulysses} spacecraft provided plasma and magnetic field data until June 2009.

\citet{Gosling94} first described the high-latitude ICMEs as a distinct class of events.
They showed that a forward-reverse shock pair is associated with each event and  this is due to the over-expansion of the CME caused by its internal pressure.
The front and reversal boundary layers, representing the interaction between the cloud and the surrounding interplanetary environment, are regions that display properties of magnetic reconnections, \ie  accelerated ion flow observed within magnetic field reversal regions in the solar wind \citep{Gosling2005}.
The outer boundary layer (between the sheath and the interplanetary environment) usually displays a drop in intensity of the magnetic field, while the inner boundary layer (which separates the interaction region from the cloud body) is associated with an initial rotation of the elevation angle and an enhancement of the magnetic field.
\citet{Wei2006JGRA} have performed a statistical analysis on $70$ magnetic clouds registered by \emph{Wind} spacecraft and found that the boundary layers are a non-pressure-balanced structures.

There are ICMEs displaying all the features, while others having only a few. Along the years, different criteria have been used to identify the ICMEs or MCs.
The presence of an ICME in the interplanetary space is often accompanied by the charge states enhancement.
\citet{Reley2013} described how the spacecraft trajectory influences the registration of an event and how observational data reveal or not the presence of an MC.
They statistically found that plasma beta, CME width, and the ratio O$^{7+}$/O$^{6+}$ are significant variables that indicate the presence of an MC, and the observation of an MC strongly depends on the spacecraft trajectory.
Using the Gaussian mixture model analysis, they identified three types of events from the ICMEs' speed \emph{vs.} plasma beta correlation: (i) a statistical populations with low plasma beta and low velocities, (ii) another with high velocities and plasma beta, and (iii) a statistical population with low velocities but high plasma beta.
The high values of the charge state of elements such as O, C, and Fe, indicate the solar flares-like events origin of the ejected plasma.
\citet{Russell2003} performed the ICMEs identification from the solar wind ion measurements.
Signatures of an expanding magnetic structure were correlated with a decline in the solar wind speed, accompanied by simultaneous cool ions.
These two indicators have long been associated with the occurrence of ICMEs (\emph{e.g.} \citet{Gosling97} and references therein).
\citet{Lepri2001} considered the enhancement of Fe as the most reliable signature of an ICME. The values of Fe charge states exceeding $11$ represent an indicative for an ICME \citep{Lepri2004}.
Another threshold that must be fulfilled by the charge states is O$^{7+}$/O$^{6+}>0.8$  \citep{Neugebauer1997}, especially in MC \cite{Henke98}.
According to \citet{Ipavich1986} and \citet{Zurbuchen2006SSRv}, the ICME threshold for the Fe/O ratio is $0.25$, since (Fe/O)$_{CME}$ / (Fe/O)$_{photosphere}>5$, with (Fe/O)$_{photosphere}=0.05$.
\citet{Henke98} analyzed more ICMEs registered by \emph{Ulysses} spacecraft and found that the MC structures have an increased O$^{7+}$/O$^{6+}$ ratio compared to the non-MC ICMEs.
In contrast with the mentioned authors that considered fixed thresholds for the charge states, \citet{Richardson04} and recently \citet{Richardson2014} developed a set of criteria that introduce thresholds charge states (for O$^{7+}$/O$^{6+}$, $<\text{Q}_ \text{Fe}>$, and Fe) depending on the solar wind speed. \citet{Richardson2014} realized a comparison between the charge states for MC and non-MC ICMEs, and compiled correlation relationships between these ones and the solar wind speed.
In the method we describe here, the charge states are less important and many times increased during the sheath of an ICME.

\citet{Richardson95} discussed in detail the temperature fall under an expected temperature value, $T_{exp}$ (see definition in next section), inside the ICMEs and heliospheric current sheets (HCS), starting from \citet{Lopez1987} formula where the temperature is computed as function of the solar wind speed.
Later, \citet{Richardson2014} computed a relationship of $T_{exp}$ threshold depending on the solar wind speed for \emph{Ulysses} data.
A comprehensive review of the ICMEs signatures has been made by \citet{Neugebauer1997}, where they introduced a thermal index for ICME identification, similar to the method developed by \citet{Richardson95}.

The problem of CMEs boundaries detection is crucial for space weather forecast, as well as for the phenomena understanding.
For that, not only one in-situ signature is sufficient to detect an event, but two or more criteria seem to be necessary in the event identification.
Many authors have used the decrease under a certain threshold of the ratio between the proton temperature and an expected computed temperature as a primordial criterion for the ICMEs identification, and considered later the beta depletion.
\citet{Wang2005jgr} performed a multispacecraft analyse of the radial evolution of the ICME registered between 0.3 and 5.4 AU using the temperature decrease as a primary criterion for the events identification. They found that the temperature is a quantity that decreases slowly with the distance from the Sun, compared with other physical quantities.
\citet{Du2010} extended the list of events detected by \citet{Wang2005jgr}, using the same criterion of identification and observations provided by \emph{Ulysses}.
\citet{Ebert2009JGRA} characterized the latitudinal and radial variations of the solar wind.
They found no significative variations in latitude, but very important ones in the radial direction. They have extended the ICME list compiled by Gosling, Reisenfeld and Forsyth (\url{http://swoops.lanl.gov/cme_list.html}), using the pitch angle distribution measurements from \emph{Ulysses} spacecraft and the magnetic field components to identify the ICMEs presence. Moreover, they used a combined set of signatures as lower than expected proton temperature, low plasma beta and high alpha particles density rated to the proton particles density. They accounted that an event is ICME if at least two criteria are fulfilled.
An additional number of events observed by \emph{Ulysses} was found by \citet{Richardson2014}, who compiled a list with 279 events.

As \citet{Gosling97} stated, the ICMEs identification ``is still something of an art'', and this because of the multiple signatures to be taken into account, but not all of them are always present for each event.
For this reason we paid attention to another tool that can help us understanding different ICME events, in order to detect the most appropriate boundaries and also to discriminate between pure or merged events.
The aim of this paper is to identify the ICME events in the solar wind plasma measured by \emph{Ulysses} spacecraft and to detect more accurate their boundaries.
In this order, we introduce an algorithm that enables a prior automatic scan of the possible event boundaries defined by certain features.
Then, by applying other magnetic field and plasma signatures, we infer the real boundaries from those previously detected.
In section 2, we describe the method as well as the data used.
We exemplify the method application by analysing four specific events, in section 3, where substructures of the ICMEs are detected, revealing complex ejecta, and the established boundaries are verified by the minimum variance analysis (MVA) tool in the MC  cases.
In section 4, we apply our method to detect the ICMEs and their boundaries using the data registered by \emph{Ulysses} during three years, 2000, 2001  and 2002, when the spacecraft flew from negative to positive latitudes, also over the solar poles and from 1.34 AU to 4.47 AU distance from the Sun.
The analyzed period contains the years with maximum of CMEs production, after the sunspots maximum and during the solar general magnetic field polarity reversal.

\section{Data and method}
\label{s2}
In the present study we used data from three instruments onboard of \emph{Ulysses} (\url{http://ufa.esac.esa.int/ufa}): VHM (magnetometer), SWOOPS (Solar Wind Plasma Experiment)  and  SWICS (Solar Wind Ion Composition Instrument).
For interplanetary events identification we have used the magnetic field data, provided by VHM instrument and given in RTN coordinates,
where \textbf{R} is the radial direction oriented from the Sun to the satellite, \textbf{T} is the cross product of the solar rotation axis and \textbf{R}, while \textbf{N} is the cross product of \textbf{R} and \textbf{T}.
The magnetic field magnitude $B$ and the ($Br$,$Bt$,$Bn$) components used to first identify the events' boundaries  had the time resolution of 1-hour,
but for refining boundaries with the minimum variance analysis we used 1-minute time resolution data.
Information on the charge states plasma composition come from SWICS with 3-hour averaged data resolution.
The solar wind plasma parameters, \ie the speed ($v$), proton density ($Np$), proton temperature ($Tp$),
were obtained by SWOOPS instrument, with hourly averaged data resolution. This resolution for SWOOPS data with the method we will describe here is optimum for our purpose, a higher resolution giving too much local variations.

SWOOPS experiment and the instrument are described by \citet{Bame92}. The data reduction algorithm adopted by SWOOPS' Team is  described in the user guide file accompanying the data.
The proton temperature has been estimated in two ways, providing two quantities, a $Tsmall$ temperature, denoted $Ts$, and a $Tlarge$ temperature, denoted $Tl$.
$Tl$ is the total numerical integral of the distribution in 3D velocity space over all energy channels and angle bins that are statistically above the noise.
The quantity $Ts$ is estimated by summing over the angle of observations at a fixed energy, and then integrating the resulting 1D moments over all energy channels.
Both temperatures are measured in the radial direction and $Ts$ is sometime underestimated, while $Tl$ is overestimated.
Since both quantities bracket the proton temperature $Tp$ (see SWOOPS user guide), different authors computed it as an average of $Tl$ and $Ts$, \emph{e.g.} \citet{Consolini2012}, \citet{Podesta2011}, or as a classical geometric mean \citet{Cranmer2009,Breech2009}:

\be Tp \equiv (Tl\cdot Ts)^{1/2}. \label{e1} \ee
If we take $Tp$ as an average, then $Tp/Ts$ reduces to $(Tl/Ts +1)/2$.
In the geometric mean case \citet{Cranmer2009}, $Tp/Ts= \sqrt{Tl/Ts}$.
Consequently, one important ratio that we use in this article, and it is independently on the adopted formula for $Tp$, is
\beq
Ar=\sqrt{Tl/Ts}.
\label{e4}
\eeq
After analysing more events, few of them described in this article, we have remarked a systematic decrease of $Ar$ ratio at the events boundaries, so our approach is to use this variation for the events detection in a semi-automatic way. This idea was used to write an IDL program able to exactly detect the boundaries with a time precision of 0.001 from a DOY (day of the year).

One important signature for an ICME is the temperature fall. During an ICME event, the proton temperature, $Tp$, rated to the expected temperature, $T_{exp}$, should be less than 0.5. This expected temperature depends on the solar wind speed, and it was computed for distances exceeding 1 AU and speed, by \citet{Wang2005jgr}, in $10^3$ K units, as:
\beq
 T_{exp} =\left\{
\begin{array}{c}
\frac{(0.031v-5.1)^{2}}{r^{0.7}},v<500 \\
\frac{(0.51v-142)}{r^{0.7}},v\geq 500%
\end{array}%
\right.
\label{ftexp}
\eeq
A new relationship of the expected temperature  for \emph{Ulysses} spacecraft, depending on the solar wind speed and spacecraft distance $r$, was computed by \citet{Richardson2014}:
\beq
Tex=(502 \cdot v -1.26 \cdot 10^5)/r
\label{ftex}
\eeq
Inside an ICME, the relationship $Tp/Tex < 0.5$ should be fulfilled.

We propose here an algorithm to detect the ICMEs boundaries, where the various known signatures are applied in an hierarchical manner.
This algorithm implies more steps:
\begin{enumerate}
  \item
  Scan of the \textit{Ulysses} data and computation of the possible interplanetary events boundaries that are the local minima (LM) of the ratio $Ar$.
  These are the potential boundaries of an event:
      $LM \in \{Ar(i)=\sqrt{Tl(i)/Ts(i)} \mid (Ar(i) < Ar(i-1)) \vee (Ar(i) < Ar(i+1)) \vee (Ar(i) < pg), i=1,...,N \}$, where $i$ denotes each registration inside a time period, $N$ represents the number of registration,  and $pg$ is a certain threshold.
   \item Plasma beta variation analysis and detection of four boundaries (at least) for each event, as follows:
     \begin{itemize}
      \item First boundary (\emph{t1}, expressed in DOY) is at the last LM where $\beta>0.2$ before a region with $\beta<0.2$
      \item Second boundary (\emph{t2}, expressed in DOY) is at the first LM with $\beta<0.2$, after \emph{t1}
      \item Third boundary (\emph{t3}, expressed in DOY) is at the last LM with $\beta<0.2$
      \item Fourth boundary (\emph{t4}, expressed in DOY) is at the following LM with $\beta>0.2$, after \emph{t3}
     \end{itemize}
  \item Analysis of the total pressure variations for the event, which could be an indicator for the extension of the event to neighbourhood boundaries related to those established at the previous step.
  \item Analysis of the other plasma and magnetic field signatures, in order to detect the complexity of an event.
  The LMs delineate very well the self-organised regions of magnetic field visualised in the contour plots of the $RTN$ components.
  These contour plots help us to detect the possible existence of MCs and to establish the preliminary boundaries, selected among the LMs, before the  minimum variance analysis. The MVA computations confirm and refine the MC boundaries.
\end{enumerate}

The LMs are defined as the amplitude of $Ar$ to be minimum relative to the nearby values.
We have limited these LMs to be under a threshold $pg$, a value established after various trails and that could be different from an event to another. However, this threshold could be $1.1<pg<2$, depending on the event.
After considering the temperature variations, we point out that the plasma beta variations are crucial for our method: if the local minima of the temperatures ratio, $Ar$, gives the potential boundaries, the $\beta$ values enable us to choose among them the final boundaries.
Plasma beta is defined as
\beq
\beta=Np\cdot k_B\cdot Tp/[B^2/(8\cdot \pi)].
\label{fbeta}
\eeq
The threshold used for plasma beta in ICMEs is 0.2, similar value used by various authors \citep{Foullon2007SoPh244,Lepping2009}.

After the raw estimation of the boundaries, other elements are taken into account for an event analysis, like the charge states anomalies, as well as the enhanced total pressure during an event.
The total protons pressure is expressed as the sum of the kinetic and magnetic pressures:
$
Pt=Np\cdot k_B\cdot Tp+B^2/(8\cdot \pi).
$
where $Np$ represents the proton number density  and $k_B$ is the Boltzman constant.
\citet{Wei2006JGRA} considered the MC boundaries as non-pressured-balanced structure.
The total pressure will help us detect an event and its expansion in time (see the case studies). As we have noticed, after the study of many events, the 'bulbs' visually created in the graph of the plasma beta and the pressure curves, as well as that created in the graph by the $Tp/Tex$ and $Ar$, delimitate an ICME. But this remark help us only to extend or restrain the time extension of an event.
Beside the four boundaries detected with the above algorithm we could also consider additional internal boundaries selected from the computed LMs and indicated by other plasma and magnetic field signatures.
The magnetic field components smooth rotation is an indicative for MC presence and we have used this methods to verify the boundaries for the analyzed events in next section.

We point out that during the important ICMEs some quantities measured by \emph{Ulysses} saturated the sensors and gave unusable values. In these circumstances we have smoothed the data, for an interval $i=1,...N$, as follows:
\begin{eqnarray}
\label{netezire}
 if(Tl(i) \geq 1.0E+006) \quad \text{then} \quad Tl(i)=Tl(i-1)\\
 if(Ts(i) \geq 1.0E+005) \quad \text{then} \quad Ts(i)=Ts(i-1)\\
 if(Np(i) \geq 99) \quad \text{then} \quad Np(i)=Np(i-1)\\
 if(v(i) \geq 999) \quad \text{then} \quad v(i)=v(i-1).
\end{eqnarray}
This smoothing procedure gave a plateau in some zones and, consequently, intervals where $Ar$ is almost constant. The software automatically chose as LM, for that plateau, the first value and the value in the middle of the interval.

\section{Case studies}
\label{s3}
In this section we exemplify the application of the method on few particular cases.
We analyze in detail three MC events we found more spectacular, and one ICME we found as new by the proposed algorithm.
Two of the MC have been summarily analyzed in previous papers by \citet{Popescu2009} and \citet{dumi_10jun}.
The first event (10 June 2001) can be found in the list of \citet{Du2010}, the second event (24 August 2001) is in the list of \citet{Ebert2009JGRA}, while the last analyzed event (23 January 2001) is found in both lists.
All three MCs could be retrieved in the recent list of  \citet{Richardson2014}.
The last event we investigate is an ICME event found as new by this algorithm, observed by the spacecraft on 1 May 2001.
In the next subsection we explain in detail, on  four examples, how we apply the method.

As we noted in introduction, more signatures accompany an ICME and not all of them are always fulfilled.
Different authors \citep{Ebert2009JGRA,Du2010,Wang2005jgr,Richardson2014} considered their own hierarchy in their compiled list of ICMEs.
In the present article, after the LM detection, we have retained those boundaries that delimited low plasma beta (see section 2). This condition is fulfilled between $t2$ and $t3$ boundaries, while $t1$ is the first LM before $t2$ and $t4$ is the first LM after $t3$ where beta exceeds the threshold.
But a very important role is played by the total pressure variation, which is certainly correlated with the temperature falling. The increase of $Pt$ represents an indication for considering the extension (or restriction) of the boundaries to next LMs, in some cases.

\subsection{The 10 June 2001 event}
\label{s3.1}
The ICME event registered on 10 June 2001 ($DOY=161$) by \emph{Ulysses} was produced by a series of explosive events occurred in the neighborhood of the active regions NOAA 09475 and 09486 \citep{dumi_10jun}.
Using hourly VHM data, 3-hours interval data from SWICS and hourly data from SWOOPS, onboard of \emph{Ulysses} spacecraft, \citet{dumi_10jun} analyzed this event and its solar source.
At that date, the spacecraft was located at a distance of 1.35 AU from the Sun and at $(19.6^o,264.6^o)$ heliographic latitude and longitude.

\begin{figure}
\begin{center}
  \includegraphics[width=9cm,height=11cm]{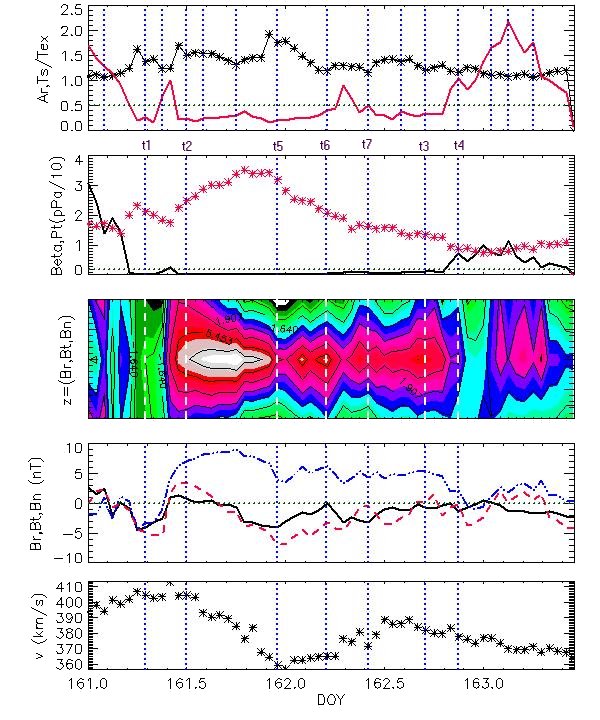}
  \caption{The event on 10 June 2001 (DOY=161).
  The vertical lines mark the boundaries: LMs for the first  panel and definitive boundaries for next panels.
  The final boundaries are: t = [ 161.292, 161.500, 161.958, 162.208, 162.417, 162.708, 162.875].
  The horizontal lines mark the applied thresholds.
  1st panel: The $Ar=\sqrt{Tl/Ts}$ quantity variations (black asterisks) and the temperatures ratio $Ts/Tex$ (solid red line) , together with the threshold $0.5$ (horizontal dotted line).
  2nd panel: Variation of the protons beta (black solid line) and total pressure (red asterisks).
  3rd panel: The contour plots of the matrix $z=(Br,Bt,Bn)$.
 4th panel: Variation of the magnetic field components, in RTN coordinates:
  $Br$ -- black, solid line, $Bt$ -- blue, dashed-dot line, $Bn$ -- red, dashed line.
  5th panel: Bulk velocity variations with time. }
  \label{161a}
\end{center}
\end{figure}

Following the algorithm described in the previous section, we have computed the possible boundaries of this event starting with the detection of the local minima of the ratio $Ar=\sqrt{Tl/Ts}$ of the temperatures registered by {\it Ulysses}.
First panel of the Fig.~\ref{161a} contains these variations of $Ar$ (black asterisks) together with the rate $Ts/Tex$ (solid red line), where the vertical lines mark the computed LM, \ie  the possible boundaries. For this event we have used $pg=1.8$.
The second panel in the figure displays the plasma beta (solid black line) and total pressure variations (red asterisks).
The vertical lines delineate the final boundaries of the event, after considering all the plasma and magnetic field characteristics, including the MVA.
 The third panel of  Fig.~\ref{161a} contains a visualisation of the magnetic field $RTN$ components \emph{vs.} time, \ie  the contour plots of the matrix $z$, for a period of few days that contains three lines and $N$ column representing the $N$ hourly registrations:
\begin{center}
$z=\left[
\begin{array}{ccc}
Br_{1} & ... & Br_{N} \\
Bt_{1} & ... & Bt_{N} \\
Bn_{1} & ... & Bn_{N}%
\end{array}%
\right] $
\end{center}
This type of visualisation, which we introduce in this article, produces very suggestive pictures of the interplanetary events, and represents an additional mode to have an estimation of the ICME boundaries. In this representation, different events are outlined as well determined, self-organised regions like 'islands', and permit an easy way to find the ICME boundaries.
Comparing the first and third panels, we remark that all LMs mark well the regions resulted from the contour plots.
The $RTN$ components of the magnetic field and the velocity variations with time are plotted in the last two panels.
Panels 3 and 4 contain the same quantities that are plotted in different manners.

This ICME displays many features of a magnetic cloud \citep{Burlaga1995,Burlaga1988}: smooth magnetic field rotates through a large angle, the strength of the magnetic field is higher than in the average solar wind, and the temperature is lower than the average solar wind.
The enhanced field strength and smooth rotation of the vertical or azimuthal component are used to identify the MC event.
The shape of these components indicates us if the event is an MC. The rotation of the magnetic field can occur in any direction on a time scale from few hours to days and can be revealed by the minimum variance analysis (MVA).
The MVA was first time applied by \citet{Sonnerup67}. \citet{KleinBurlaga82} and \citet{BothmerSchwenn98} described in detail the MVA method applied to MCs.
The directional changes of the magnetic field can be investigated with the MVA applied to the field components in the spacecraft frame. Therefore, we pursued to detect these changes of orientation for all substructures of the magnetic cloud orientations in the Cartesian solar equatorial coordinates ($X,Y,Z$), where $Bx=-Br$, $By=-Bt$, and $Bz=Bn$, \emph{i.e.} $By$ pointing to the East relative to the Sun (as described in \citet{Dumi2011}). This fact implies that RTN coordinates are rotated by 180$^\circ$ around the normal axis \citep{Mulligan2001}.
After the MVA computations, we have obtained the magnetic field components of the clouds $Bxc,Byc,Bzc$ corresponding to the maximum variance, the intermediate and minimum variance directions, respectively. The results are relevant if the ratio of the eigenvalues corresponding to the intermediate and the minimum variance is under $2$,
$\lambda_2/\lambda_3 \geq 2$,
and directional changes of the magnetic field vector exceed $30^\circ$. All these conditions are fulfilled in our computations for all three analyzed events in this paper.

The contour plots of the magnetic field (panel 3, Fig.~\ref{161a}) indicate that the studied event is complex and has more components, and more internal boundaries should be taken into account.
After the plasma and magnetic field characteristics, and the numerical results of the minimum variance analysis (MVA) examination, we have selected between the LMs the following event boundaries, expressed in days of the year ($DOY$):
$t1=161.29$, $t2=161.50$, $t3=162.708$, $t4=162.875$.
As we can see in Fig.~\ref{161a}, supplementary internal boundaries (between $t2$ and $t3$) were also considered at
$t5=161.958$, $t6=162.208$ and $t7=162.417$.
These have been selected among the LMs, after various combinations of LMs in the MVA computations and after analysing the plasma characteristics too (for confirmation of the chosen boundaries). The MVA results are discussed below.
Between $t1$ and $t2$ it exists an LM (at $t=161.375$), that normally should be considered as $t2$,  indicated also by the contour plots of the magnetic insulation. But, we selected the next LM as $t2$, after the MVA computations.

\begin{figure}[h!]
\begin{center}
  \includegraphics[width=0.8\textwidth]{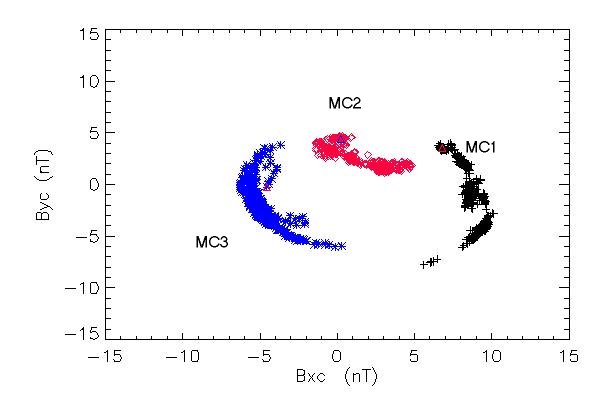}
  \caption{The  MVA results for the event registered on 10 June 2001  (DOY=161). See the text for details. }
  \label{161mva2}
\end{center}
\end{figure}

After MVA computations with different combinations of boundaries, we can conclude that the 10 June 2001 event is a complex one and has the following morphology: there are three magnetic clouds (denoted M1, M2 and M3). Their boundaries are:
\begin{itemize}
  \item MC1 lasts from  $t2=161.500$ to $t5= 161.958 $
  \item  MC2 lasts from  $t5=161.958$ to $t6=  162.208 $
  \item  MC3 lasts from  $t7=162.417$ to $t3=  162.708 $.
\end{itemize}
Between $t6$ and $t7$ there are the reverse shock of MC2 and the forward shock of MC3.

We have applied the MVA using 1-minute averaged data registered by VHM instrument and have obtained the following azimuthal ($\phi$) and elevation rotation angles ($\theta$) for the three axis, \ie  for the maximum, intermediate, and minimum variance directions \citep{BothmerSchwenn98}, for all the MCs.
The results are:
\begin{center}
$ \phi1= [     171.74,      180.00,      263.21] $ \\
$ \theta1= [     41.81,      90.00,      89.89] $ \\
$ \phi2=  [    253.89,      180.00,      173.68] $ \\
$ \theta2=  [    89.86,      90.00,      46.15] $ \\
$ \phi3=   [   223.41,      179.79,      106.68] $ \\
$ \theta3=  [    5.16,      90.00,     -90.00] $ \\
\end{center}

Fig.~\ref{161mva2} shows our  minimum variance analysis computation results, plotted in the intermediate-maximum variance plane (YOX) for all clouds.
The triangle symbol on each MC plot  marks the start point of the cloud rotation at the spacecraft encounter. The first MC has a negative rotation, while the other two MCs display a rotation in the positive sense. According with \citet{BothmerSchwenn98}, MC1 has a left-handed helicity, MC2 has a right-handed helicity, as well as MC3. The different helicities of the clouds could indicates their different solar sources (flares from two distinct active regions).

\subsection{The 24 August 2001 event}
\label{s3.2}
Another MC event was observed by \emph{Ulysses} on 24 August 2001, \ie  $DOY=236$
\citep{Popescu2009,Ebert2009JGRA,Richardson2014},
when the spacecraft was located at 1.68 AU and $285.5^o$ longitude, $67.3^o$ latitude.
The track back to the Sun of this ICME , using a graphical method described in \citet{Dumi2011}, gave a raw estimation of the solar source as being a CME occurred on $DOY=230.25$. The CDAW catalogue (\url{http:\\cdaw.gsfc.nasa.gov\CME_list}) indicates, as possible solar counterparts, two CMEs in squall (a series of CMEs coming from the same solar source, see \citet{Dumi2008}) occurred at the polar angle $PA\sim 337$ deg. These are the results of a double polar sigmoid filament eruption in more steps, on 15 August 2001 ($DOY=227$).
A CME was registered by LASCO/SOHO at 1:31 UT, with a speed 477 km s$^{-1}$, another CME occurred at 2:54 UT, with 370 km s$^{-1}$ and a CME at 10:33 UT, with v=311 km s$^{-1}$. A linear propagation model of the CMEs indicates the arrival of these events at \emph{Ulysses} in the proper time. It remains the problem of the CMEs' speeds that are much smaller than the speed measured in-situ by  \emph{Ulysses} for the ICME (530 km s$^{-1}$), if we do not suppose that the ICME was accelerated by the ambient solar wind.
Another candidate for the solar source is the halo CME occurred on 19 August 2001 ($DOY=231$), 6:06 UT, in or nearby the active region NOAA 09575, with the speed v=556 km s$^{-1}$. This halo CME is due to a flare.

\begin{figure}[h!]
\begin{center}
  \includegraphics[width=9cm,height=10cm]{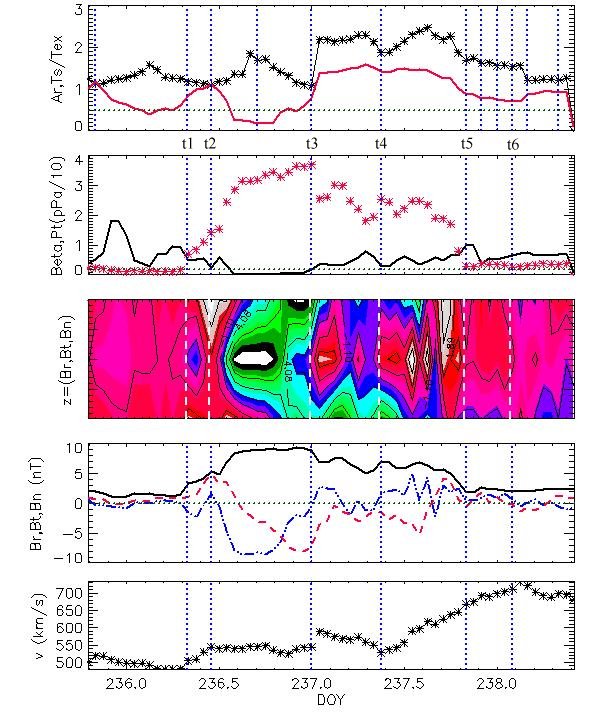}
  \caption{The event on 24 August 2001 (DOY=236) -- caption similar to Fig.~\ref{161a}.
  The final events' boundaries are
   t = [236.330, 236.458, 237.000, 237.375, 237.833,238.083]}.
  \label{236a}
\end{center}
\end{figure}

Fig.~\ref{236a} displays the magnetic field and plasma characteristics of the event, where the procedure of the boundaries detection is similar to that applied at the previous section, according to the described algorithm. We have used $pg=2$ for these LMs computations.

After examining the magnetic field contour plots and performing the MVA computation we have found the following boundaries of the event:
 $t1=236.330$, $t2=236.458$, $t3=237.0$, and $t4=237.375$.
This event is also a mixed one: it is a MC close followed by a stream interaction region (SIR). The SIR is not preceded by a heliospheric current sheet (HCS) since it is at very high latitude.
The forward shock is at $t1$  and MC sheath lasts before $t2$. The MC core extends between $t2$ and $t3$. At $t3$ is the reverse shock of the MC and between $t3$ and $t4$ is the trailing region of the MC.
Looking at the magnetic field contour plots and the velocity curve shape, we see that $t4$ marks the forward shock of the SIR, while $t6=237.833$ marks the SIR reverse shock. The end of the SIR at $t6$ is sustained also by the velocity shape, but considering the pressure values we could suspect this reverse shock as being at $t5=237.375$ .

\begin{figure}[h!]
\begin{center}
  \includegraphics[width=0.6\textwidth]{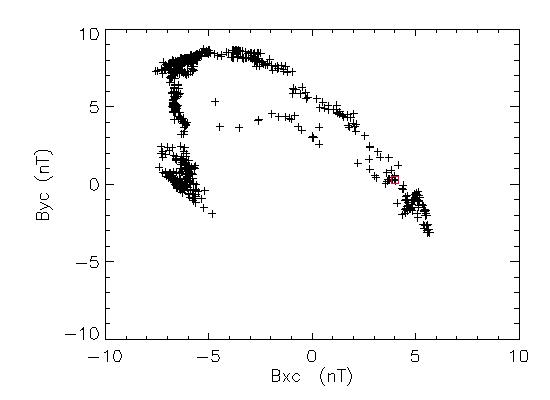}
  \caption{The 24 August 2001 event (DOY=236) --  MVA computations plot for the MC. } \label{236mva2}
\end{center}
\end{figure}

We have performed the MVA for the MC core, between $t2$ and $t3$ (781 points), and the result is displayed in Fig. \ref{236mva2}.
The azimuthal and elevation angles of the three axis obtained in this analysis are:\\
$\phi=[      143.291  ,    112.784 ,     180.0]$,
$\theta=   [   44.653,  -89.931, 90.0]$.
This MC has a left-handed helicity.

The SIR following the MC could be an acceleration factor for the MC event, explaining the difference between the initial CME's speeds registered by \emph{SOHO} spacecraft and the ICME event's speed registered to the \emph{Ulysses} spacecraft.
An interesting region is between $t3$ and $t4$, where the magnetic field contour plots indicate the existence of two small 'islands' and in their front the temperature and pressure have a severe discontinuity - the reverse shock of the MC. After the moment $t3$, the pressure decreases compared with that of MC, but it is still high. This MC trailing region could be a zone of interactions, but we suspect that there is a small remanent ejecta embedded in the SIR, which could be the signature of the solar source (the CMEs in squall).
The existence of ICMEs embedded in SIRs are reported in the literature also by other authors \citep{Rodriguez2005}.

\subsection{The 23 January 2001 event}
\label{s3.3}
Two halo CMEs, registered by CDAW catalogue, came from two X-flares on 20 January 2001: first CME occurred at 19:31 UT, with a speed of 839 km s$^{-1}$ and the second at 21:30 UT, with 1507 km s$^{-1}$. These flares were observed in the active region NOAA 09313.
Considering a linear propagation, the corresponding interplanetary counterparts of those CMEs should hit the spacecraft, located at 1.88 AU, $236.2^0$ longitude and $-67.6^o$ latitude, on 23 January and 22 January, respectively. A well-defined MC was observed by \emph{Ulysses} starting with 23 January 2001  \citep{Ebert2009JGRA,Du2010,Richardson2014}.

\begin{figure}
\begin{center}
  \includegraphics[width=9cm,height=10cm]{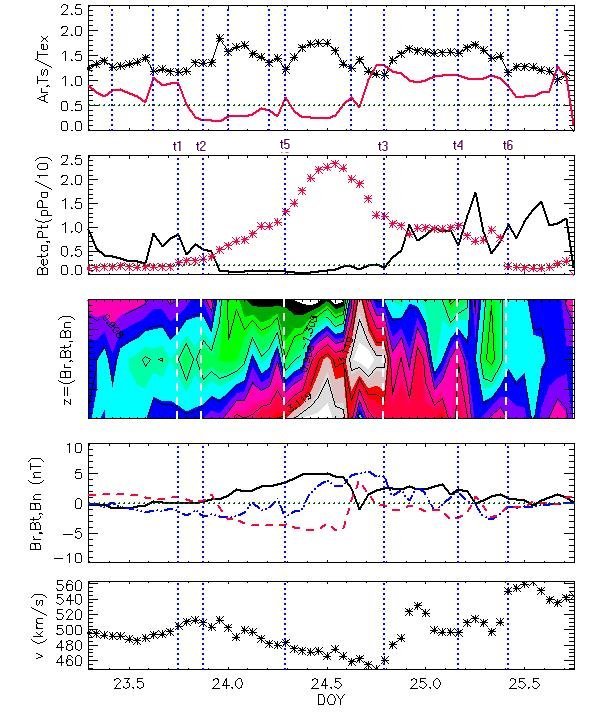}
  \caption{The event on 23 January 2001 (DOY=23): caption similar to Fig.~\ref{161a}.
  The final boundaries \hspace{20cm} are t = [ 23.750, 23.875, 24.291, 24.791, 25.041, 25.416].}
  \label{23a}
\end{center}
\end{figure}

\begin{figure}
\begin{center}
  \includegraphics[width=6cm,height=6cm]{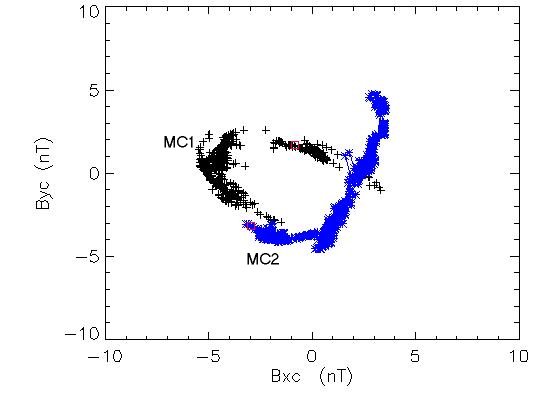}
   \includegraphics[width=6.5cm,height=6.5cm]{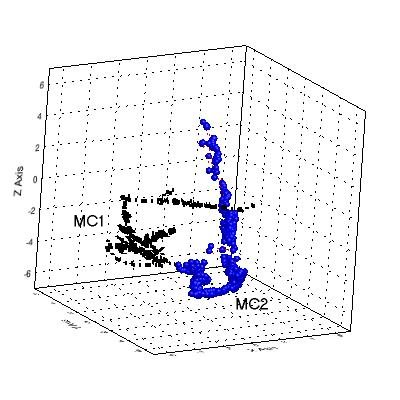}
  \caption{The 23 January 2001 (DOY=23) event --  MVA computations plot for the MC. } \label{23mva2}
\end{center}
\end{figure}

Fig.~\ref{23a} displays the characteristics of this event: the temperatures ratio $Ar$ used for the potential boundaries detection (the LMs computation), the temperature ratio $Ts/Tex$, variations of the plasma beta  and total pressure, the magnetic field components (with two visualisations) and the velocity variations.
Using the algorithm proposed in this article, we have obtained the following first boundaries, after the plasma beta values evolutions.
The contour plots in Fig.~\ref{23a} suggest the existence of at least two structures. After the MVA computations for different cases we have established the following boundaries:
$t1=23.750$, $t2= 23.875$,  $t3=  24.791$, $t4= 25.041$, with an intermediate boundary at $t5=  24.291$ representing a boundary between two magnetic clouds.
Consequently, MC1 lasts from $t2$ to $t5$ and MC2 from $t5$ to $t3$. The forward shock is at $t1$, and the reverse shock at $t3$, with the trailing region extended until $t4$. The two MC are merged and the second one seems to be twisted, as the 3D representation of the MVA computations reveals in Fig.~\ref{23mva2}.
The double-core MC is followed by a fast stream that ends at $t6=  25.416 $.

We have performed MVA computations for different combinations of the boundaries and considering the coalescence of the two structures.
The MVA results confirm the existence of an internal boundary and two magnetic sub-clouds.
The best results (Fig.~\ref{23mva2}) were obtained for  computations performed between $t2$ and $t5$ (MC1 -- 600 points)  and between $t5$ and $t3$ (MC2 -- 720 points), giving the azimuthal and elevation angles for the maximum, intermediate, and minimum variance directions:\\
%
$\phi1=    [  240.80,      129.51 ,     180.00]$,
$\theta1=    [  21.44,     -90.00  ,    90.00]$, \\
$\phi12=    [ 208.14,      155.24 ,     155.24]$,
$\theta2=    [  22.95,      90.00 ,     90.00]$.

\subsection{The 1 May 2001 event }
\label{s3.4}
The algorithm we have settled up allows us to detect also small events, not spectacular and often masked by neighbourhood events, but that obey to our criteria. One of such event was observed by \emph{Ulysses} at the beginning of May 2001, when the spacecraft was positioned at $-12.26$ deg. latitude and $259$ deg. longitude, and a distance of 1.37 AU from the Sun.
This event was not classified by other authors before, and it is part of a complex series composed by a MC observed on 29 April 2001 (DOY 119),  listed by \citet{Richardson2014}, followed by a small ICME, detected and described by us in this article, and a heliospheric current sheet (HCS) that ends on 3 May 2001 (DOY 123).

\begin{figure}
\begin{center}
  \includegraphics[width=10cm,height=11.75cm]{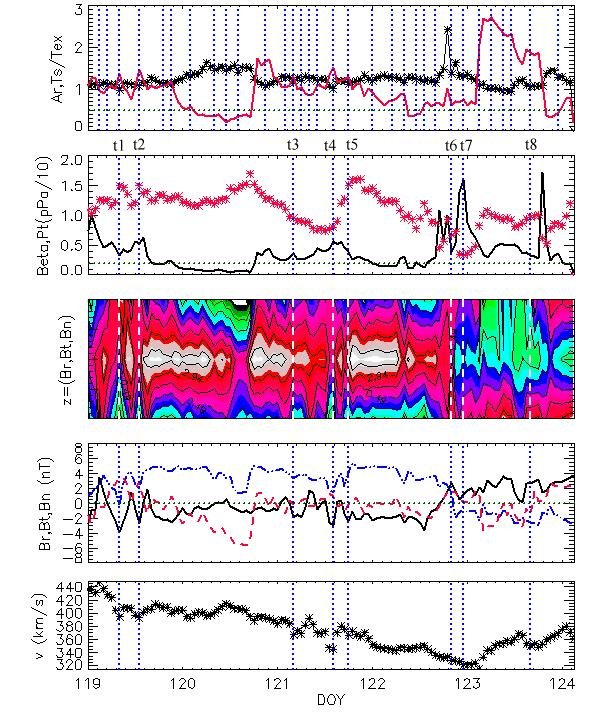}
  \caption{The complex interplanetary events registered between 29 April - 3 May 2001 - caption similar to Fig.~1.
Final boundaries are t = [ 119.333, 119.542, 121.167, 121.583, 121.750, 122.833, 122.958, 123.667].}
  \label{119a}
\end{center}
\end{figure}

Fig.~\ref{119a} displays the characteristics of these interplanetary events. After the LMs computation, we have selected the following boundaries:
the MC has the forward shock at $t1=119.333$, the core between $t2=119.542$, $t3=121.167$, and the trailing edge ending at $t4=121.583$. The core of the cloud is sectioned in two parts, at $t=120.458$ , as we can see on the contour plots in the Fig.~\ref{119a}, but this division is not sustained by the other parameters as plasma beta, pressure and the total magnetic field intensity variations. Between $t3$ and $t4$ we can consider the presence of the trailing edge of the MC. This MC is followed by a complex ejecta that has the shock at $t4$ (more visible in the velocity) and the sheath between $t4$ and $t5=121.750$. The ICME lasts between $t5$ and $t6=122.833$.
All these events are accompanied by a SIR, existing from $t7=122.958$ to $t8=123.667$, and marked by the sudden and high increase of the plasma beta, temperature and pressure. This SIR is preceded by a HCS at $t6$ (HCS being marked by the magnetic field signs reversal).
We remind that \emph{Ulysses} crossed the ecliptic at that time.
Looking on the coronal EIT/SOHO observations two days before 1 May 2002, indeed, a small coronal hole faced the Earth at small latitudes.

%
\begin{figure}
\begin{center}
 \includegraphics[width=10cm,height=8.5cm]{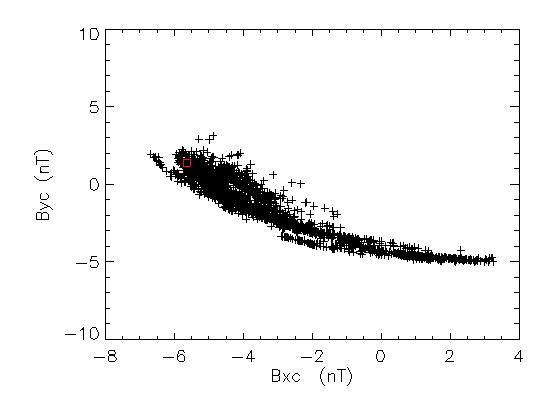}
  \caption{The MVA plot of the MC registered on 29 April 2001  (DOY=119). } \label{121mva2}
\end{center}
\end{figure}

We have performed the minimum variance analysis, on  2340 points, for the MC registered between $t2$ and $t3$.
The result is displayed in Fig.~\ref{121mva2}, and the  obtained azimuthal and elevation angles are:
\vspace{7pt}
\begin{center}
$\phi=[192.67 ,     180.00,      123.12]$;
$\theta= [ 5.99,     90.00,     -90.00]$.
\end{center}

\section{ICMEs list and algorithm validation}
\label{s4}
In order to verify our method on many other interplanetary events, we have applied the above described algorithm and compiled a ICMEs list for a period of three important years of the solar cycle 23, namely 2000, 2001 and 2002.
We have found 63 events, listed in the Tables 1,2 and 3. These events could be retrieved in the lists compiled by \citet{Ebert2009JGRA}, \citet{Du2010}, and \citet{Richardson2014}.
By following the criteria we introduced in this article, we found 9 new events too.
The  table contains data as follows:
each event line has four boundaries $t1,t2,t3,t4$ together with the corresponding velocity and proton beta.
The $t1$ and $t4$ boundaries are roughly defined for $\beta>0.2$, while $t2$ and $t3$ are established at the LM with $\beta<0.2$.
There are few exceptions, namely when the aspect of the magnetic field or total pressure ($Pt$) indicates a more extended event -- then we have considered as boundaries the corresponding other LM. It is also the case of the 10 June 2001 event.

\citet{Reinard2008ApJ} suggested the ICMEs have structures consisting of a core (or multiple cores) and an envelope. This morphology is consistent with the observed CMEs structure near the Sun \citep{Gopal2006}.


\begin{center}
\begin{landscape}
\begin{table}
{\footnotesize
\centering
\caption{The ICMEs list 2000} \label{table1}
\begin{tabular}{|l|l|l|l|l|l|l|l|l|l|l|l|l|}
\hline\hline
No& t1 &v1 &$\beta1$ &t2 &v2 &$\beta2$ &t3 &v3&$\beta3$ &t4&v4  & $\beta4$\\
&  (DOY)  & (km s$^{-1}$) &&(DOY) &(km s$^{-1}$) & &(DOY) &(km s$^{-1}$)  &&(DOY) & (km s$^{-1}$) & \\
\hline\hline
1 & 17.83 & 421.2 & 0.89 & 18.04 & 417.6 & 0.24 & 18.58 & 397.7 & 0.28 & 18.88 & 400.2 & 0.62   \\
2 & 39.13 & 548.6 & 2.56 & 39.46 & 539.8 & 0.17 & 39.92 & 545.9 & 0.2 & 40 & 530.3 & 0.78   \\
3 & 40.67 & 539.3 & 0.26 & 41.13 & 528.8 & 0.16 & 42.08 & 524.3 & 0.15 & 42.25 & 501.3 & 0.21   \\
4 & 91.29 & 394.7 & 1.18 & 91.46 & 386.2 & 0.5 & 92.25 & 401.8 & 0.16 & 92.38 & 401.1 & 1.09   \\
5 & 103.67 & 431.4 & 0.34 & 103.79 & 426.6 & 0.12 & 105.17 & 438.1 & 0.28 & 105.25 & 443.8 & 0.4   \\
6 & 121.17 & 514.6 & 0.71 & 121.46 & 486 & 0.04 & 123.38 & 410.9 & 0.05 & 123.71 & 392.1 & 0.3   \\
7 & 133.71 & 465.2 & 0.39 & 133.88 & 465 & 0.09 & 135.29 & 443 & 0.11 & 135.5 & 457.2 & 0.41   \\
8 & 147.63 & 385.3 & 0.81 & 147.79 & 376.4 & 0.21 & 148.71 & 366.7 & 0.08 & 149.29 & 377 & 1.3   \\
9 & 177 & 352.6 & 0.41 & 177.13 & 358.1 & 0.09 & 178 & 342.9 & 0.18 & 178.13 & 341.3 & 0.2   \\
10 & 178.13 & 341.3 & 0.2 & 178.25 & 341.7 & 0.15 & 179.88 & 320.4 & 0.21 & 180.04 & 341 & 0.45   \\
11 & 223.42 & 481.8 & 0.48 & 223.67 & 472.7 & 0.09 & 225.25 & 427.9 & 0.18 & 225.5 & 434.4 & 0.25  \\
12 & 256.88 & 407.1 & 0.97 & 257.04 & 390.9 & 0.02 & 257.88 & 401.6 & 0.94 & 258.13 & 403.7 & 0.78   \\
13 & 291.96 & 384 & 0.8 & 292.17 & 378.8 & 0.03 & 293.17 & 381.7 & 0.08 & 293.54 & 376.7 & 0.28  \\
14 & 294.25 & 433.4 & 0.83 & 294.46 & 404.3 & 0.42 & 294.88 & 413.8 & 0.22 & 295.21 & 456.7 & 2.3  \\
15 & 313.92 & 383.7 & 0.68 & 314.04 & 362.5 & 0.23 & 314.21 & 360.5 & 0.08 & 314.46 & 410.5 & 0.24  \\
16 & 341.33 & 445.8 & 0.42 & 341.58 & 423.1 & 0.14 & 344.21 & 338.5 & 0.2 & 344.33 & 343.1 & 0.29  \\
\hline\hline
\end{tabular}
}
\end{table}
\end{landscape}

\begin{sidewaystable}
\centering
\caption{The ICMEs list 2001} \label{table2}
{\footnotesize
\begin{tabular}{|l|l|l|l|l|l|l|l|l|l|l|l|l|}
\hline\hline
No& t1 &v1 &$\beta1$ &t2 &v2 &$\beta2$ &t3 &v3&$\beta3$ &t4&v4  & $\beta4$\\
&  (DOY)  & (km s$^{-1}$) &&(DOY) &(km s$^{-1}$) & &(DOY) &(km s$^{-1}$)  &&(DOY) & (km s$^{-1}$) & \\
\hline\hline
17 & 23.75 & 504.8 & 0.85 & 23.88 & 509.6 & 0.53 & 24.79 & 460.8 & 0.15 & 25.04 & 499.6 & 0.99  \\
18 & 32.54 & 469.8 & 0.64 & 32.79 & 447.6 & 0.15 & 33.25 & 430.4 & 0.09 & 33.58 & 426 & 1.23  \\
19 & 55.54 & 327.9 & 0.23 & 55.63 & 327.4 & 0.11 & 56.33 & 302.4 & 0.18 & 56.67 & 298.6 & 0.9  \\
20 & 78.04 & 423.9 & 0.54 & 78.13 & 400.9 & 0.6 & 78.75 & 358.8 & 0.27 & 78.88 & 354 & 0.33  \\
21 & 79.88 & 359.1 & 1.03 & 80.04 & 359.9 & 0.24 & 80.88 & 346.4 & 0.24 & 81.33 & 355.8 & 0.48  \\
22 & 90.67 & 366.8 & 1.25 & 90.83 & 363.3 & 0.31 & 91.42 & 346.6 & 0.17 & 91.75 & 372.4 & 0.38  \\
23 & 92 & 373.9 & 0.29 & 92.38 & 327.5 & 0.14 & 93.08 & 304.9 & 0.06 & 93.33 & 387.3 & 0.25   \\
24 & 100.75 & 575.9 & 0.52 & 100.92 & 602.1 & 0.06 & 102.88 & 492.5 & 0.16 & 103 & 503.8 & 0.22  \\
25 & 108.25 & 349.3 & 0.21 & 108.54 & 372.1 & 0.17 & 109.46 & 471.4 & 0.16 & 109.67 & 524.6 & 0.63   \\
26 & 109.83 & 520.9 & 0.52 & 110 & 537.5 & 0.17 & 113.83 & 460.4 & 0.15 & 114.08 & 449.1 & 0.25   \\
27 & 118.5 & 459.6 & 0.34 & 118.58 & 447.5 & 0.18 & 118.92 & 435.5 & 0.4 & 119 & 436 & 0.75  \\
28 & 119.54 & 392.8 & 0.53 & 119.79 & 408.4 & 0.17 & 120.58 & 406.4 & 0.06 & 120.88 & 392.7 & 0.41  \\
29 & 121.58 & 345 & 0.56 & 121.75 & 366.8 & 0.4 & 122.67 & 336.1 & 0.3 & 122.83 & 328.5 & 0.38   \\
30 & 130.83 & 602.9 & 0.3 & 131.04 & 617.3 & 0.13 & 132.5 & 486.3 & 0.18 & 132.67 & 492.7 & 0.25  \\
31 & 139.13 & 425.8 & 0.41 & 139.29 & 433.9 & 0.1 & 140.71 & 362.5 & 0.18 & 140.83 & 368.6 & 0.25  \\
32 & 156.25 & 462.4 & 0.21 & 156.46 & 515.2 & 0.09 & 156.75 & 526.2 & 0.14 & 156.92 & 505.3 & 0.24  \\
33 & 161.38 & 403.6 & 0.13 & 161.5 & 404.5 & 0.01 & 162.71 & 381.9 & 0.11 & 162.88 & 377.8 & 0.72  \\
34 & 185.63 & 300.1 & 0.3 & 185.75 & 298.3 & 0.19 & 186.46 & 295.3 & 0.12 & 186.83 & 308.6 & 0.48   \\
35 & 187.21 & 303.8 & 0.48 & 187.42 & 303 & 0.13 & 188.25 & 300.8 & 0.05 & 188.54 & 294.4 & 0.45  \\
36 & 204.25 & 407.6 & 0.86 & 205.29 & 379.4 & 0.22 & 206.08 & 357.7 & 0.26 & 206.17 & 353.5 & 0.52  \\
37 & 235.83 & 519.4 & 0.5 & 236.46 & 543.3 & 0.24 & 237 & 544.4 & 0.19 & 237.38 & 528.7 & 0.29  \\
38 & 271.46 & 769.3 & 0.47 & 272.17 & 687.3 & 0.23 & 272.29 & 669.3 & 0.18 & 272.75 & 671.4 & 0.66   \\
39 & 318.25 & 632.6 & 1.1 & 318.42 & 652.9 & 0.94 & 319.58 & 613 & 0.12 & 319.67 & 618 & 0.6   \\
40 & 331.21 & 881.6 & 0.9 & 331.54 & 806.6 & 0.35 & 331.75 & 759.9 & 0.12 & 332.17 & 752.7 & 0.72   \\
\hline\hline
\end{tabular}
}
\end{sidewaystable}

\begin{landscape}
\begin{table}
\centering
\caption{The ICMEs list 2002} \label{table3}
{\footnotesize
\begin{tabular}{|l|l|l|l|l|l|l|l|l|l|l|l|l|}
\hline\hline
No& t1 &v1 &$\beta1$ &t2 &v2 &$\beta2$ &t3 &v3&$\beta3$ &t4&v4  & $\beta4$\\
&  (DOY)  & (km s$^{-1}$) &&(DOY) &(km s$^{-1}$) & &(DOY) &(km s$^{-1}$)  &&(DOY) & (km s$^{-1}$) & \\
\hline\hline
41 & 2.71 & 562.1 & 0.31 & 2.83 & 564.2 & 0.15 & 3.54 & 571.8 & 0.12 & 3.63 & 554.1 & 0.46   \\
42 & 17.92 & 516 & 0.66 & 18.04 & 531.5 & 0.37 & 19.29 & 484.5 & 0.24 & 19.54 & 470.2 & 1.4  \\
43 & 31.46 & 419.8 & 0.81 & 31.63 & 414.7 & 0.23 & 32.17 & 404.7 & 0.21 & 32.33 & 416.7 & 0.25   \\
44 & 32.46 & 412 & 0.47 & 32.63 & 396.3 & 0.08 & 32.88 & 405.1 & 0.1 & 33.04 & 430.6 & 0.65   \\
45 & 42.63 & 540.1 & 0.72 & 42.83 & 546.7 & 0.13 & 43.88 & 523 & 0.29 & 44.17 & 526.3 & 1.15 \\
46 & 73.29 & 503.6 & 0.48 & 73.42 & 504 & 0.18 & 73.96 & 484.7 & 0.27 & 74.08 & 496.4 & 0.27  \\
47 & 84.25 & 489 & 2.23 & 84.42 & 484.9 & 0.07 & 84.71 & 478.9 & 0.09 & 85.04 & 496.8 & 0.36   \\
48 & 96.5 & 427.1 & 2.09 & 96.75 & 415.2 & 0.52 & 97.58 & 400 & 0.12 & 97.67 & 386.7 & 1.16  \\
49 & 125.04 & 390.7 & 0.3 & 125.17 & 396.5 & 0.15 & 129.42 & 322.1 & 0.17 & 129.67 & 436.3 & 1  \\
50 & 155 & 470.9 & 1.01 & 155.08 & 479.2 & 0.19 & 155.96 & 717.8 & 0.82 & 156.08 & 720.1 & 0.74   \\
51 & 156.54 & 726.5 & 1.13 & 156.67 & 722.2 & 0.11 & 157.08 & 707.2 & 0.26 & 157.25 & 708.5 & 1.07  \\
52 & 164.5 & 775.1 & 1.15 & 164.63 & 835.9 & 0.78 & 169.17 & 564.9 & 0.02 & 169.96 & 529.2 & 0.21   \\
53 & 197.92 & 448.3 & 0.49 & 198.04 & 486 & 0.16 & 199.5 & 532 & 0.05 & 199.71 & 528.7 & 0.24  \\
54 & 210.29 & 673.6 & 0.36 & 210.5 & 693.1 & 0.23 & 211.13 & 703.7 & 1.6 & 211.79 & 728 & 0.58   \\
55 & 224.33 & 450.5 & 0.25 & 224.67 & 426.4 & 0.11 & 226.71 & 405.8 & 0.15 & 226.92 & 424.4 & 0.31   \\
56 & 227 & 425.2 & 0.39 & 227.13 & 424.1 & 0.18 & 227.88 & 396.5 & 0.07 & 228 & 378.7 & 0.46   \\
57 & 274.33 & 398.9 & 0.22 & 274.5 & 395.3 & 0.1 & 275.5 & 374.4 & 0.14 & 275.83 & 368.5 & 0.4   \\
58 & 276.33 & 397.4 & 0.56 & 276.42 & 401.3 & 0.05 & 276.79 & 399.6 & 0.07 & 276.92 & 391.8 & 0.48   \\
59 & 291.96 & 460.7 & 0.72 & 292.54 & 430.6 & 0.17 & 293.04 & 420.5 & 0.19 & 293.17 & 415 & 0.25   \\
60 & 300.46 & 366 & 0.31 & 300.63 & 368.3 & 0.13 & 302.13 & 360.8 & 0.09 & 302.42 & 358.8 & 0.55  \\
61 & 308.17 & 501.8 & 0.76 & 309.13 & 514.7 & 0.1 & 309.58 & 497.9 & 0.02 & 309.83 & 492.1 & 0.96   \\
62 & 316.46 & 417.5 & 0.35 & 316.92 & 488.6 & 0.26 & 317.75 & 460.3 & 0.18 & 317.83 & 469.3 & 0.64 \\
63 & 325.33 & 389 & 2.37 & 325.63 & 385.8 & 1.63 & 327.71 & 371.1 & 0.18 & 327.79 & 369.6 & 0.28 \\
\hline\hline
\end{tabular}
}
\end{table}
\end{landscape}


\end{center}


\begin{figure}[h!]
\begin{center}
  \includegraphics[width=0.7\textwidth]{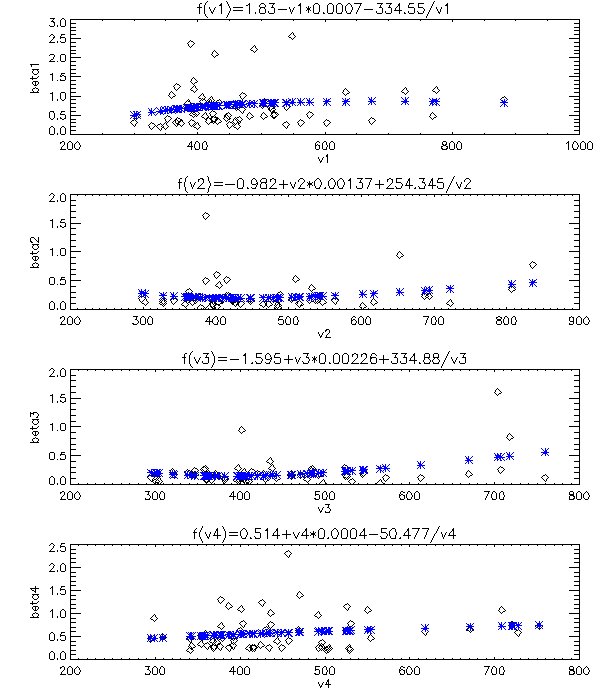}
  \caption{Velocities \emph{vs.} plasma beta for each category of boundaries: observed (diamonds) and computed after fitting (asterisks). } \label{st}
\end{center}
\end{figure}

We have computed the sizes of each ICME listed in the tables, expressed in astronomical unities (AU), by considering that the core extends between $t2$ and $t3$, while the surrounding region (the envelope) expands between $t1$ and $t2$, and between $t3$ and $t4$.
We have obtained the mean core of $0.35$ AU, while the whole event (core and envelope) has a mean size of $0.48$ AU.
The average size of the leading region, between $t1$ and $t2$, is $0.06$ AU, and average size of the trailing region, between $t3$ and $t4$, is $0.07$ AU.

We remind here the algorithm we have applied to detect the moments $t1$,$t2$,$t3$,$t4$. At the first step we have computed the LMs of the ratio $Ar=\sqrt{Tl/Ts}$, \ie  the local minima mathematically defined for the data series within a certain period of time. These LMs are only possible boundaries since we search through them the true boundaries, after applying the classical ICMEs signatures. The contour plots of the magnetic field are very helpful in the boundaries selection too.
We choose $t2$ and $t3$ as the first and the last boundaries where $\beta<0.2$, while $t1$ and $t4$ are those boundaries with $\beta>0.2$ positioned at the left, respectively at the right, of $t2$ and $t3$.

After establishing these main boundaries, we examined the other plasma and magnetic field signatures in order to decide if an event is extended or not, or it reveals internal sub-structures.
Fig.~\ref{st} displays the data for each computed boundary of the events listed in the tables, velocities \emph{vs.} plasma beta (diamonds).
The asterisk traces display the fitted curve by the formula  $f(v)=a+b\cdot v+c/v$, where $v$ takes values from 300 to 900 km s$^{-1}$.
This formula was found after various trials. We obtained for each boundary a fitted curve that states a plasma $\beta$ dependence on the speed values. This fact could be explained by the formula of $\beta$ (see formula \ref{fbeta}), where the proton temperature is incorporated, and the expected temperatures were usually obtained as function of speed (see formulas \ref{ftexp} and \ref{ftex}).
We remark that the curves fitted for the first and last boundary, where shocks usually occur, especially for MC, have the same concave shape, while for the second and third boundary we obtained convex curves.

\section{Summary and discussions}
\label{s5}
The prediction of CMEs' arrival time is an important goal for heliospheric researches and space weather alerts. This prediction is linked to the propagation mode of the CME into the interplanetary space, but also to the accuracy of the ICME boundaries detection. Usually, these boundaries could be established by the analysis of magnetic field components curves' shape, or adding information about the plasma $\beta$, temperature, density, speed and charge states variation.
We proposed here a new method to detect the interplanetary event boundaries using the local minima of the \emph{Ulysses'} temperatures ratio and the variations of plasma beta as first criterion to delimit the ICMEs boundaries.
Our method, in this form, is applicable for \emph{Ulysses} spacecraft.
Both temperatures measured by \emph{Ulysses}, $Ts$ and $Tl$, are in the radial direction and each one could be used as proton temperature. On the other hand, it exists a systematic variation of their ratio with local minima at the interplanetary events' boundaries. Consequently, we deduce that the SWOOPS instrument could have the best precision of the measurements at the boundaries of the events and a maximum variation inside an event or an interplanetary structure.
We note that the local minima of  $Ar$  appear at the boundaries between different plasma flow regimes, before or after shocks, as well as outlining the plasma magnetic insulation.
These local minima could be an indicative of the isotropization of the field-aligned streams (or counterstreams), frequently observed in the boundary regions after the forward and reverse shocks \citep{Skoug2000, Steed2011, Lazar2014}.

The algorithm we have applied here allowed us to compute the boundaries of an event following few steps.

(1) The LMs computation using an IDL code. These LMs delineate the contour plots 'islands' of the magnetic field components.

(2) Selection of those two LMs that border a region with $\beta<0.2$ and other two LMs enclose the first two but correspond to $\beta>0.2$.
Generally, the  selected region for an ICME event fit well to regions with the temperature failure.

(3) The analysis of the total pressure variation  and magnetic field signatures could be an indicative to extend or to limit the boundaries of an event.

(4) In the individual cases analysis we could find that intermediate boundaries are necessary (belonging also to the computed LM), if merged or complex events exist.

(5) The minimum variance analysis validates the selected boundaries for the MCs.

We have analyzed the morphology of three known MCs and a complex of events containing one new found ICME.
Our method allowed us to better understand the morphology of the complex or hybrid events.

We have compiled the boundaries for the ICMEs occurred during the maximum of solar CME production period, \ie  the years 2000, 2001 and 2002 and found events listed also by other authors, but also few new small ICMEs.
We have fitted the values of plasma beta found at the boundaries as function of the corresponding velocities, by a formula of $f(v)=a+b*\cdot v+c/v$ type.

The ICMEs detection is semi-automatic since the first step is to scan the possible boundaries using a computer software. This software operation detects the possible boundaries. Then a human intervention keeps the right boundaries after analyzing the plasma beta, pressure and magnetic field contour plots.
This scan covered an interval of three years and was performed with a 10-15 days period.
The scan of data allowed us to find new small events, if we look to the other authors' lists. There are also few other events that do not belong to our list even if there are in the other authors’ lists and this because the second criteria ($\beta<0.2$) was not accomplished during the whole period of the event.

The method described here is useful for the identification of the ICME observed by \emph{Ulysses} and their boundaries, since this spacecraft measures two proton temperatures and the automatic detection relies on their rate.
We notice as original contribution of this paper the introduction of magnetic field contour plots visualisation that reveals the plasma insulations and allows in this way the boundaries detection.
This type of visualisation could be used in the case of other spacecraft, \ie \emph{ACE} for instance.
But in this case the procedure becomes a visually method for establishing the boundaries and the precision is under $0.001$, less than in the \emph{Ulysses} case.

\vspace{-12pt}
\begin{acknowledgements}
We thank to SWOOPS' and SWICS Teams for their work to delivery the data to the scientific community.
We are especially grateful to Dr. Ruth Skoug and Dr. Bruce Goldstein for their useful clarifications on SWOOPS data.
We acknowledge the National Space Science Data Center and the Principal Investigator,
Dr. A. Balogh  of Imperial College, London, UK, for the VHM data.
The CME catalog is generated and maintained at the CDAW Data Center by NASA and The
Catholic University of America in cooperation with the Naval Research Laboratory, and use data from SOHO.
\emph{Ulysses} and SOHO are projects of international cooperation between ESA and NASA.
We also thanks Dr. Marian Lazar (Ruhr-Universitt Bochum) for useful discussions on the topic.
\end{acknowledgements}
\makeatletter
\def\@biblabel#1{}
\makeatother

\received{\it 1 October 2015}

\end{document}